\documentclass[journal]{IEEEtran}
\ifCLASSINFOpdf
\else
\fi
\usepackage{amsthm,amsmath,amssymb}
\usepackage{bm}
\usepackage{mathrsfs}

\usepackage[ruled]{algorithm2e}
\usepackage{graphicx}
\usepackage{booktabs} 
\usepackage{float} 
\usepackage{booktabs}
\usepackage{multirow}
\usepackage{array}
\usepackage{stfloats}
\usepackage{color}
\usepackage{cite}
\usepackage[dvipsnames, svgnames, x11names, table,xcdraw]{xcolor}
\usepackage{courier}
\usepackage[numbers,sort&compress]{natbib}

\begin{document}
%
\title{Learning First-Order Rules with Relational Path Contrast for Inductive Relation Reasoning}
%
%
%

\author{Yudai~Pan,
        Jun~Liu, Lingling~Zhang, Xin~Hu, Tianzhe~Zhao and Qika~Lin
\thanks{Yudai Pan is with the School of Computer Science and Technology, Xi’an Jiaotong University,
 China e-mail:
	pyd418@foxmail.com}
\thanks{Jun Liu, Lingling Zhang, Xin Hu, Tianzhe Zhao and Qika Lin are with the School of Computer Science and 
Technology, Xi’an Jiaotong 
University}
\thanks{This work has been submitted to the IEEE for possible publication. Copyright may be transferred without notice, 
after which this version may no longer be accessible.}}

%
%

\markboth{Journal of \LaTeX\ Class Files,~Vol.~14, No.~8, August~2015}%
{Shell \MakeLowercase{\textit{et al.}}: Bare Demo of IEEEtran.cls for IEEE Journals}
%



\maketitle

\begin{abstract}
Relation reasoning in knowledge graphs (KGs) aims at predicting missing relations in incomplete triples, whereas the 
dominant paradigm is learning the embeddings of relations and entities, which is limited to a transductive setting and 
has restriction on processing unseen entities in an inductive situation. Previous inductive methods are scalable and 
consume less resource. They utilize the structure of entities and triples in subgraphs to own inductive ability. 
However, in order to obtain better reasoning results, the model should acquire entity-independent relational semantics 
in latent rules and 
solve the deficient supervision caused by scarcity of rules in subgraphs. To address these issues, we propose a 
novel graph convolutional network (GCN)-based approach for interpretable inductive reasoning with relational path 
contrast, named RPC-IR. RPC-IR firstly extracts relational paths between two entities and learns representations of 
them, and then innovatively introduces a contrastive strategy by constructing positive 
and negative relational paths. A joint training strategy considering both supervised and contrastive information is 
also proposed. Comprehensive experiments on three inductive datasets show that RPC-IR achieves outstanding performance 
comparing with the latest inductive reasoning methods and could explicitly represent logical rules for interpretability.
\end{abstract}

\begin{IEEEkeywords}
Knowledge graph, Inductive learning, Rule learning
\end{IEEEkeywords}

%
\IEEEpeerreviewmaketitle

\section{Introduction} \label{section_intro}
%
%
%
%
\IEEEPARstart{K}{nowledge} graphs (KGs) store plenty of knowledge by a collection of triples consisting of entities and 
relations. 
With the development of KGs including Freebase \cite{DBLP:conf/sigmod/BollackerEPST08}, 
YAGO \cite{DBLP:conf/www/SuchanekKW07}, DBpedia \cite{DBLP:conf/semweb/AuerBKLCI07} and Wikipedia 
\cite{DBLP:conf/www/Vrandecic12}, the dominant methods have been proposed to learn representations by mapping 
relations and entities into low-dimension vectors or matrices (i.e. embeddings), such as 
translation-based models \cite{DBLP:conf/nips/BordesUGWY13, DBLP:conf/aaai/WangZFC14}, or bilinear models 
\cite{DBLP:journals/corr/YangYHGD14a, DBLP:conf/icml/NickelTK11}.
These methods can be used to predict relations in incomplete triples reasoning.   

\begin{figure}[t]
	\centering
	\includegraphics[scale=0.72]{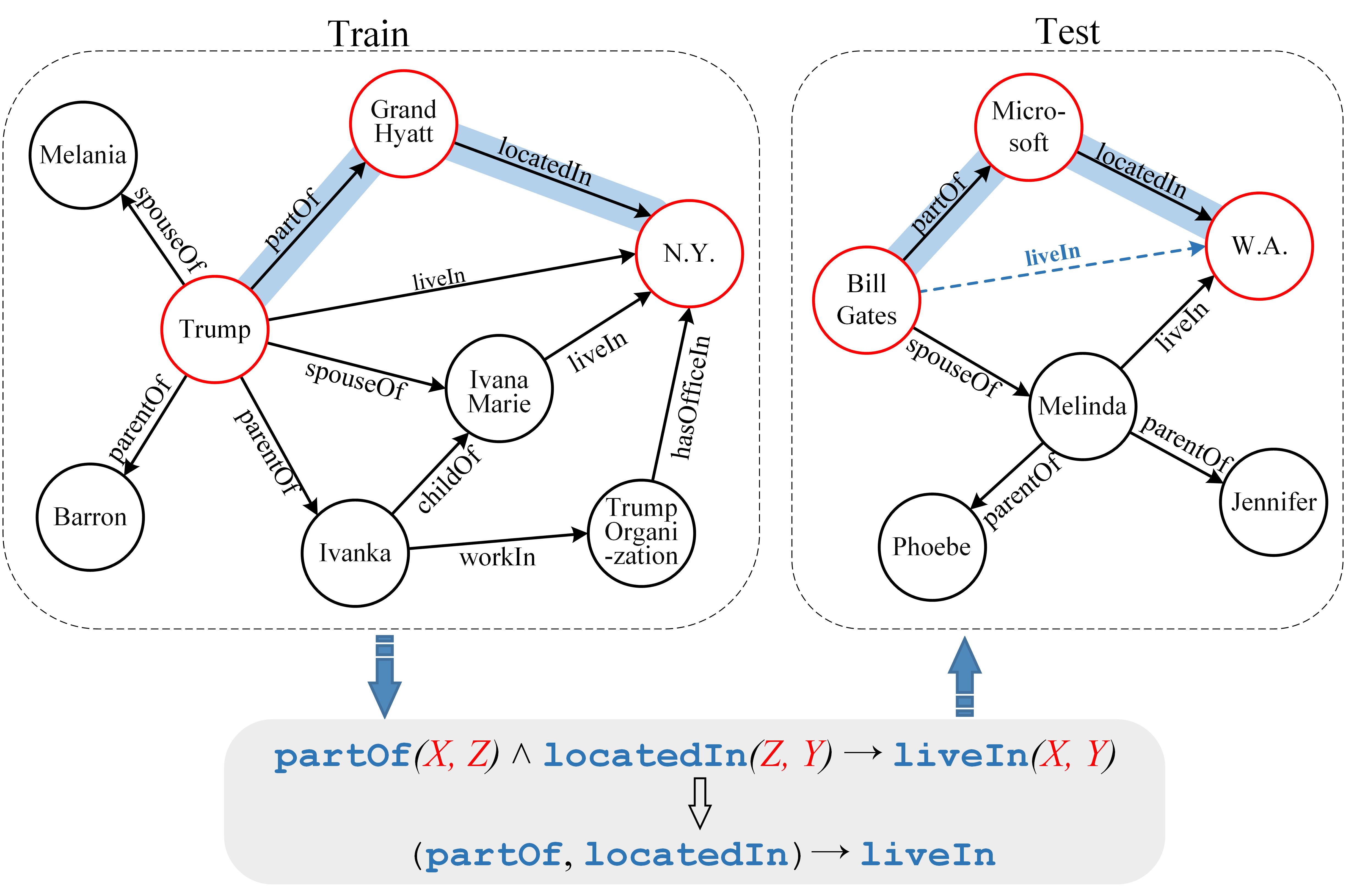}
	\caption{An example for inductive reasoning.}
	\label{fig:intro}
\end{figure}

However, the reasoning task implemented by above methods assumes for a \emph{transductive} setting, which means the 
entities are fixed during training and testing. According to the open-world assumption, there are new data 
distributions in the test set meaning unseen entities are waiting for testing, referring to the \emph{inductive} 
setting.
For example, considering the scenario in Fig. \ref{fig:intro}, entities in the train set and test set have no 
intersection, so the previous transductive methods will not accurately predict the relation denoted as the blue dotted 
arrow without retraining the whole model.
Thus, we focus on models owning inductive ability, which can handle unseen entities in reasoning by mining latent rules 
during training process, for example, 
\begin{equation}\label{equ:rule1}
\begin{small}
\texttt{partOf}(X, Z) \land \texttt{locatedIn}(Z, Y) \to \texttt{liveIn}(X, Y),
\end{small}
\end{equation}
where $X, Y, Z$ are variables in the logical rule. The arrow points to the head of Rule \eqref{equ:rule1} and the 
rest is the body. From Fig. \ref{fig:intro}, the relation \texttt{liveIn} between entities \texttt{Bill Gates} and 
\texttt{W.A.} in the
test subgraph can be inferred by Rule \eqref{equ:rule1} without retraining the model. 
Existing models of inductive reasoning, for example \cite{DBLP:conf/icml/TeruDH20}, majorly take advantage of the 
topological structure of entities and triples in subgraphs, which have an inductive capability 
and consume less resources. However, there remain two challenges of mining latent rules in inductive reasoning. 

Firstly, inductive reasoning has a problem that the test set owns unseen entities during training, which requires 
entity-independent relational semantics contained in latent rules. 
The topological structure of entities and triples is difficult to capture semantics of relation sequences in rules 
which are more critical for entity independence.
For example, Rule \eqref{equ:rule1} indicates that entities \texttt{Trump}, \texttt{Grand Hyatt} and \texttt{N.Y.} are 
generalized to variables $X, Y, Z$ marked in red in Fig. \ref{fig:intro}, so the entities are less important than 
connection of relations (\texttt{partOf}, \texttt{locatedIn}) for predicting the relation \texttt{liveIn} during 
testing as shown in the gray box.


Secondly, subgraphs own the inductive capability \cite{DBLP:conf/icml/TeruDH20}, but the scarcity of rules in a single 
subgraph leads to deficient supervision in inductive learning. Previous statistical methods 
\cite{DBLP:conf/ijcai/MeilickeCRS19, 
	DBLP:conf/www/GalarragaTHS13} indicate that in a KG owning $m$ kinds of relations, the complexity of candidate 
	rules' number within length $n$ is $O(m^n)$, whereas a subgraph only contains a few latent rules. 
For example in the train subgraph of Fig. \ref{fig:intro}, there are actually four rules whose length within 3 from 
entity \texttt{Trump} to \texttt{N.Y.}, but at least $8^3=512$ candidate rules, which is far more than the actual 
rules. As a note, we treat the reasoning paths as rules like thick blue paths in Fig. \ref{fig:intro}. More detailed 
statistics of rules in subgraphs from different datasets are shown in TABLE \ref{table:intro}.
It is unlikely to obtain all the supervised information of candidate rules in a subgraph, which would reduce the 
performance of the model.

%
\begin{table}[t]
	\centering
	\caption{Statistics of average rules from all subgraphs with the max rule length as 3 respectively in a KG. There 
	are four versions in each dataset, whose details are indicated in TABLE \ref{table:1}}
	\renewcommand\arraystretch{1.3}
	\label{table:intro}
	\begin{tabular}{l|llll}
		\hline \hline
		Dataset   & v1     & v2    & v3     & v4   \\ 
		\hline
		WN18RR
		
		& 1.46      & 1.47      & 1.43    & 1.48         \\ \hline
		FB15K-237  		
		
		& 3.13     & 9.31    & 16.76    & 25.27        \\ \hline
		NELL-995 		
		
		& 4.68     & 15.40    & 20.95    & 24.22        \\ \hline
		\hline
	\end{tabular}
\end{table}

Considering to solve the above challenges, we propose an interpretable approach based on \textbf{R}elational 
\textbf{P}ath \textbf{C}ontrast for \textbf{I}nductive \textbf{R}easoning named RPC-IR. In order to acquire 
entity-independent relational semantics, RPC-IR extracts relational paths within a preset length like (\texttt{liveIn}, 
\texttt{locatedIn}) in Fig. \ref{fig:intro}, obtaining 
representations without variables and entities. To address the deficient supervision of rules in a single subgraph, we 
propose a contrastive strategy, which is a kind of self-supervised learning methods by constructing positive and 
negative relational paths. 
After that, RPC-IR obtains representations of positive and negative relational paths using a graph convolution network 
(GCN), and proposes a joint training strategy combining the supervised and self-supervised information. In the end, 
RPC-IR obtains the structure of first-order rules like Rule (1) by relational paths in a single subgraph. The learned 
rules with confidences can explain the reasoning process in KGs.


Our main contributions can be summarized into three folds:
\begin{itemize}
	\item An inductive reasoning approach RPC-IR is proposed. We utilize relational paths to represent 
	rules in subgraphs, and design a path representation method to capture the entity-independent information of rules.
	\item We innovatively devise a contrastive strategy to solve deficient supervision of rules in subgraphs. We 
	firstly employ contrastive learning into inductive reasoning and rule learning tasks.
	\item Experiments of the relation prediction task on three inductive datasets verify the superiority and 
	effectiveness of RPC-IR. It achieves outstanding performance comparing with latest inductive reasoning methods and 
	explicitly shows the first-order rules for interpretability. 
\end{itemize}

The rest of this paper is organized as follows: Section \ref{section_rela} surveys previous work about inductive 
learning in KGs and contrastive learning. Section \ref{section_method} comprehensively illustrates the proposed RPC-IR. 
Section \ref{section_exp} demonstrates the effectiveness of RPC-IR by extensive experiments. In section 
\ref{section_con}, we conclude the whole paper and put forward the future 
work.

\section{Related Work} \label{section_rela}
We have surveyed the previous related work about inductive learning in KGs and contrastive learning respectively. 

\subsection{Inductive Learning in Knowledge Graphs}
Inductive learning in KGs requires models own generalization for handling the unseen nodes, which could be 
divided into two aspects: rule-based and graph-based.



\subsubsection{Rule-based}
Previous rule-based methods induce inherent rules from KGs according to statistical patterns. AMIE 
\cite{DBLP:conf/www/GalarragaTHS13}, RuleN \cite{DBLP:conf/semweb/MeilickeFWRGS18}, AnyBURL 
\cite{DBLP:conf/ijcai/MeilickeCRS19} and RLvLR \cite{DBLP:conf/ijcai/OmranWW18} mine entity-independent rules by 
enumerating all the candidates and select rules by preset thresholds. However, these statistical methods are 
limited to the time complexity and scalability. Different with these, some models are proposed to induce rules in a 
differentiable pattern, which means to train the model and learn rules by gradient descent in KGs. Yang et al. 
\cite{DBLP:conf/nips/YangYC17} firstly proposed a differentiable model Neural-LP to learn rules, obtaining the 
structure and confidence of rules simultaneously by a neural controller system. 
Sadeghian et al. \cite{DBLP:conf/nips/SadeghianADW19} utilized the bidirectional recurrent neural network (RNN) to 
capture the backward and forward information about the order of atoms in rules and learn rules with variable lengths. 
Wang et al.\cite{DBLP:conf/iclr/WangSDK20} proposed Neural-Num-LP to extend Neural-LP by learning numerical rules and 
Qu \cite{DBLP:journals/corr/abs-2010-04029} extended Neural-LP by an EM-based algorithm to learn high-quality rules. 
The above differentiable methods are based on a framework named TensorLog \cite{DBLP:journals/corr/Cohen16b} to 
represent the triples using matrices, whose dimension is the number of entities, so the space complexity would be high. 
From the descriptions of previous work, rule-based inductive methods in statistical and differentiable pattern will 
cost 
enormous time and space resources respectively.

\subsubsection{Graph-based}
To solve the scalablity and complexity issue, some graph-based methods are proposed for the inductive setting by 
extracting subgraphs. 
Teru et al. \cite{DBLP:conf/icml/TeruDH20} proposed GraIL to extract subgraphs from KGs and implement the inductive 
learning by a graph neural network (GNN) with an edge attention mechanism. Mai et al. \cite{DBLP:conf/aaai/MAI} 
proposed CoMPILE to strengthen the message interactions between edges and entities through a communicative kernel, and 
enable a sufficient flow of relation information. Compared with rule-based inductive methods, graph-based inductive 
methods are 
more scalable. Distinguished with the above methods, RPC-IR not only utilizes the graph structure, but also captures 
the entity-independent information by relational paths and solve the deficient supervision of latent rules in 
subgraphs. In addition, it obtains interpretability in KG reasoning.

%

\subsection{Contrastive Learning}
Contrastive learning, which is a framework of self-supervised learning, consists of methods that obtain representations 
by learning to encode samples similar or different. Contrastive learning is utilized in the natural language, computer 
vision and graph domains.
As a work in the natural language domain, Oord et al. proposed a contrastive method named CPC 
\cite{DBLP:journals/corr/Oord} to get context latent representations by predicting future information, using a 
probabilistic contrastive loss. 
In the field of computer vision, He et al. \cite{DBLP:conf/cvpr/He0WXG20} proposed a contrastive learning framework 
MoCo for visual representation, which builds large and consistent dictionaries for unsupervised learning with a 
momentum contrastive loss. Another contrastive method for visual representation is SimCLR 
\cite{DBLP:conf/icml/ChenK0H20}, which declares that the composition of data augmentation is crucial for contrastive 
tasks, and illustrates that contrastive learning benefits from larger sizes and more training epochs. 
In the graph domain, Velickovic \cite{DBLP:conf/iclr/VelickovicFHLBH19} proposed deep graph infomax to contrast the 
patch representations and corresponding high-level summary of graphs. Kipf et al. \cite{DBLP:conf/iclr/KipfPW20} 
introduced C-SWMs, utilizing a novel object-level contrastive strategy for representation in environments with 
compositional structure modeled by GNNs. 

These methods utilize contrastive learning for representation on text, images and graphs etc., improving the 
effectiveness on different downstream tasks. For solving the deficient supervision of rules in subgraphs, we 
innovatively utilize the contrastive strategy into the inductive reasoning task.

\begin{figure*}[htbp]
	\centering
	\includegraphics[scale=0.75]{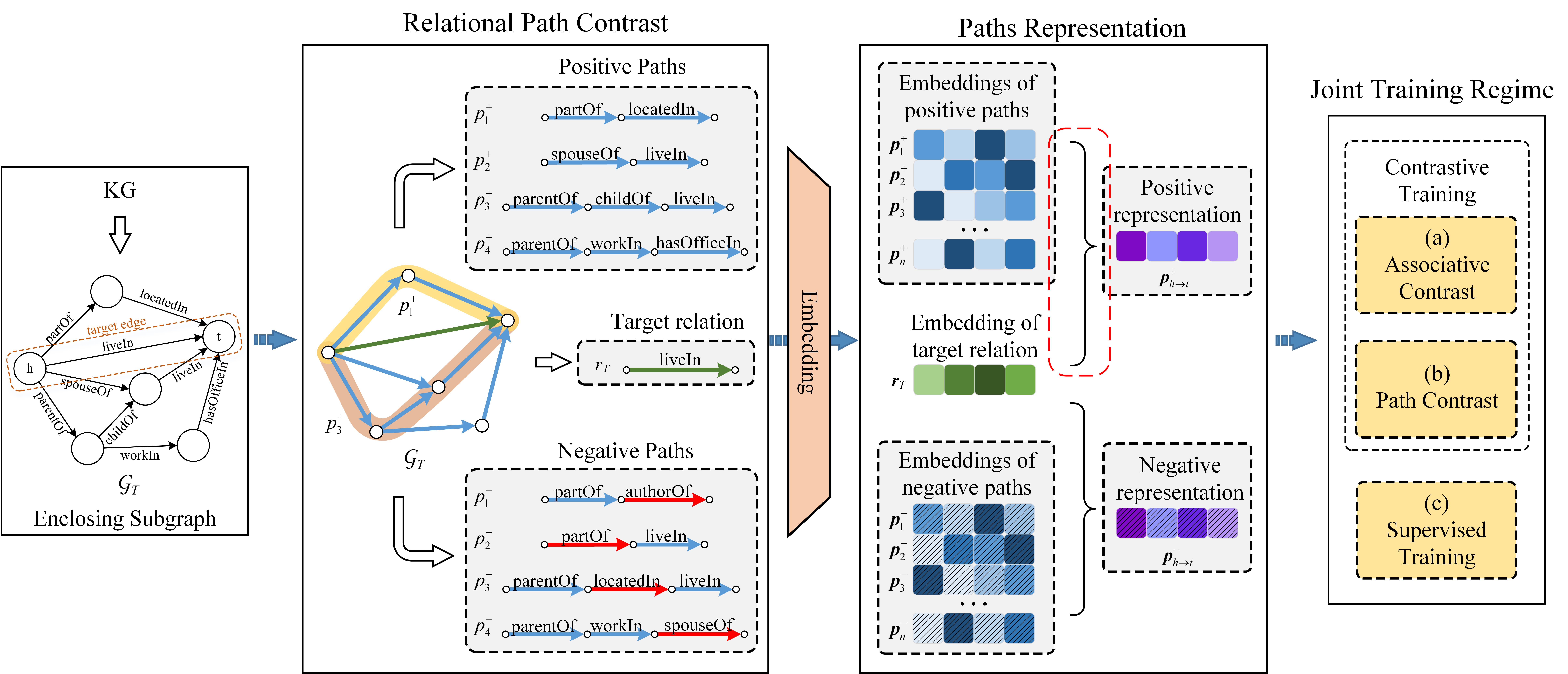}
	\caption{The framework our RPC-IR.}
	\label{fig:1}
\end{figure*}

\section{Methodology} \label{section_method}

This section comprehensively illustrates the inductive reasoning task and our proposed approach RPC-IR. 

\subsection{Task Definition and Overview of RPC-IR}
Inductive relation reasoning in KGs is to make relation prediction on unseen entities. 
A target triple $e_{\scriptscriptstyle T}$ is known as a triple $(h, r_{\scriptscriptstyle T}, t)$ in the train KG $G 
\left \langle 
R,E\right 
\rangle$, in 
which 
$h$ and $t$ refer to the head entity and tail entity, and $r_{\scriptscriptstyle T}$ is the target relation. $R$ and 
$E$ are sets of relations and entities in $G$.
Relation prediction in a fully-inductive setting intends to quantify the score of target triple $e_{\scriptscriptstyle 
T}$ and predict the relation between two unseen entities $h'$ and $t'$ in a testing KG $G' \left \langle R', E'\right 
\rangle$, where $R' \subseteq R$ and $E' \cap E= \varnothing $. 
In our proposed method, the set of relational paths $\mathcal{P}_{h \to t}$ in the enclosing subgraph of target 
triple $e_{\scriptscriptstyle T}$ are extracted and used to score $e_{\scriptscriptstyle T}$. 
The RPC-IR can be divided into three steps: 1) extracting paths from the enclosing subgraph of the target triple, and 
producing positive and negative samples of relational paths for contrast; 2) obtaining representations of positive and 
negative samples using a GCN; 3) scoring the target triple with the subgraph and relational paths, and training the 
model by a joint training strategy. 
After these steps, learned rules are attained by relational paths with their confidences. 
The demonstration of three steps are shown in the following subsections with the help of Fig. \ref{fig:1}.

\subsection{Initialization and Contrast Construction}
Firstly in this step, we extract subgraphs from $G$ and obtain features of nodes. Then we generate contrastive samples 
by constructing positive and negative relational paths. The details are as follows.  

\textbf{Node Features.} 
We extract the enclosing subgraph $\mathcal{G}_{T}$ based on the target triple 
$e_{\scriptscriptstyle T}$ from $G$, and implement 
the double radius vertex labeling scheme to entities in the subgraph.  The node $i$ around target triple $(h, 
r_{\scriptscriptstyle T}, t)$ 
is in the intersection 
of k-hop undirected neighborhoods of $h$ and $t$. Following 
\cite{DBLP:conf/nips/ZhangC18}, the node is labeled as $(d(i, h), d(i, t))$, in which $d$ is the shortest topological 
distance between two entities. 
The label of node $i$ is denoted as [$\text{one-hot}(d(i, h));\text{one-hot}(d(i, t))]$ $\in \mathbb{R}^{(2k+2)}$ to 
indicate the node feature, where $[\cdot ; \cdot]$ refers to the concatenation operation. 


\textbf{Contrastive Relational Paths Generation.}
After that, relational paths need extracting from the subgraph. 
In order to select all the topological relational paths of subgraph $\mathcal{G}_{T}$, we use 
breadth first search (BFS) algorithm for extracting every path whose length is no longer than $L_{max}$ from $h$ to 
$t$. The set of extracted paths is denoted as $\mathcal{P}_{h \to t}$. 
For instance in Fig. \ref{fig:1}, if $L_{max}$ is set as 3, then the algorithm would select 4 relational paths from the 
extracted subgraph 
$\mathcal{G}_{T}$. 
Moreover, 
we design a strategy to generate contrastive relational paths in $\mathcal{G}_{T}$ by constructing positive and 
negative samples. We consider the target relation $r_{\scriptscriptstyle T}$ 
as the original instance and extracted relational paths as the positive path samples. 
As for the negative samples, they are constructed to distinguish semantics with the original instance and 
positive samples, so we randomly replace a part of every relational path, and avoid it appearing in the set of positive 
samples. 
For example, if the extracted path is $(\texttt{partOf}, \texttt{locatedIn})$, the negative path would be 
$(\texttt{partOf}, \texttt{authorOf})$, in which the replaced relation is denoted as the red arrow in Fig. \ref{fig:1}. 
In following descriptions, 
the original 
relation in the subgraph is denoted as $r_{\scriptscriptstyle T}$, the $i$-th positive and negative path samples are 
denoted as $p^{+}_i$ 
and $p^{-}_i$ 
respectively. The sets of positive and negative paths are correspondingly indicated as $\mathcal{P}_{h \to t}^{+}$ and 
$\mathcal{P}_{h \to t}^{-}$. 

\subsection{Paths Representation}
The second step of RPC-IR is to get representations of relational paths in the subgraph $\mathcal{G}_{T}$. We obtain 
embeddings of entities and relations using a GCN, and design a strategy for paths representation. The details are in 
the following.

\textbf{Subgraph Embedding.} 
We implement a GCN \cite{DBLP:conf/esws/SchlichtkrullKB18} for obtaining embeddings of entities and relations in the 
KG. 
The propagation process for calculating the forward-pass update is defined as:
\begin{equation}\label{equ:graph1}
\bm{z}^{(k+1)}_i = \text{ReLU}(\sum_{r \in R} \sum_{j \in \mathcal{N}_i^r} \alpha_{i,r} \textbf{W}_r^{(k)} 
\bm{z}_j^{(k)} + \textbf{W}_{self}^{(k)} 
\bm{z}_i^{(k)}),
\end{equation}
where $\bm{z}^{(k+1)}_i$ denotes the embedding of node $i$ in the $(k+1)$-th layer. $\mathcal{N}_i^r$ denotes the set 
of neighbors of entity $i$ connected by relation $r$. $\textbf{W}_r^{(k)}$ and $\textbf{W}_{self}^{(k)}$ 
refer to the transformation matrices for propagating messages from layer $k$ to $k+1$. $\alpha_{i,r}$ is the edge 
attention weight corresponding to 
the edge connected via $r$, which is obtained following \cite{DBLP:conf/icml/TeruDH20}:
\begin{align}
\label{equ:graph2} \bm{y}_{i,r} & = \sigma_1(\textbf{W}_1[\bm{z}^{(k)}_i ; \bm{z}^{(k)}_j ; \bm{r} ; 
\bm{r}_{\scriptscriptstyle T}] + \bm{b}_1), \\
\label{equ:graph3} \alpha_{i,r} & = \sigma_2(\textbf{W}_2 \bm{y}_{i,r} + \bm{b}_2).
\end{align} 
$\bm{r}$ and $\bm{r}_{\scriptscriptstyle T}$ indicate the embeddings 
of relation $r$ and target relation $r_{\scriptscriptstyle T}$ respectively. $\sigma_1$ and $\sigma_2$ are activation 
functions, such as 
ReLU($\cdot$) or 
Sigmoid($\cdot$).

\begin{figure}[t]
	\centering
	\includegraphics[scale=0.68]{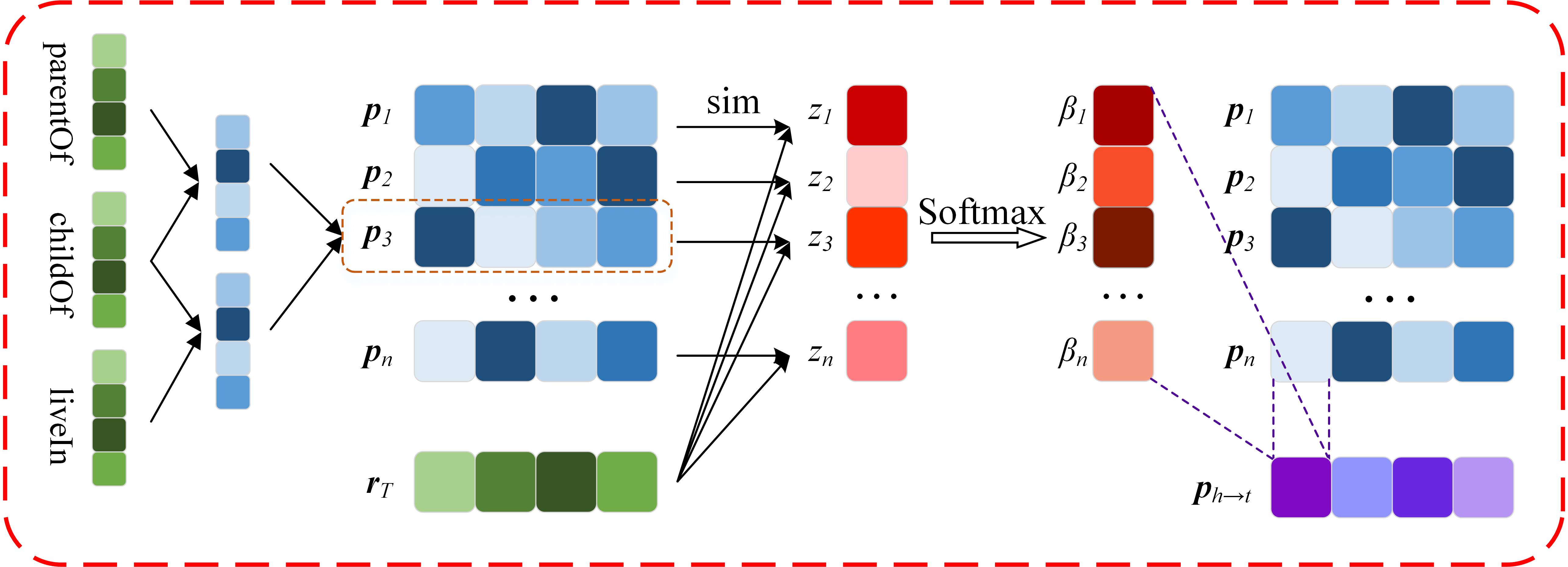}
	\caption{The process for paths representation.}
	\label{fig:2}
\end{figure}

\textbf{Paths Representation.} 
We design a strategy to obtain representations of relational paths in the subgraph $\mathcal{G}_{T}$, which is shown in 
the red dotted block of Fig. \ref{fig:1} and Fig. \ref{fig:2}. In the paths representation step, we use the embeddings 
of the enclosing subgraph $\mathcal{G}_{T}$ with entities and relations in it.  

Inspired by a rule mining work \cite{DBLP:journals/corr/YangYHGD14a}, the relational paths are used to represent the 
inference process by rules. We calculate the semantic similarity between the target relation $r_{\scriptscriptstyle T}$ 
and the relational 
path $p_i \in \mathcal{P}_{h \to t}$, for $r_{\scriptscriptstyle T}$ and $p_i$ connect the same $h$ and $t$. Then, we 
utilize an aggregation 
function $\psi$ to obtain the representation:  
\begin{equation}\label{equ:path4}
\bm{p}_{h \to t} = \psi(\{\bm{p}_i: p_i \in \mathcal{P}_{h \to t}\}).
\end{equation}
As illustrated in Fig. 
\ref{fig:2}, $p_1, p_2, \cdots, p_n$ are $n$ paths in $\mathcal{G}_{T}$, 
then the representation of the paths is given by:
\begin{equation}\label{equ:path1}
\bm{p}_{h \to t} = \sum^n_{i=1}\beta_i \bm{p}_i
\end{equation}
in which $\beta_i$ is the attention weight between the path $p_i$ and $r_{\scriptscriptstyle T}$. $\bm{p}_i$ is the 
representation of path $p_i$, and we add representations of relations that constructing $p_i$ to represent $\bm{p}_i$, 
which is implemented by the continuous bag-of-words (CBOW) algorithm. For an alternate strategy, the path 
representation can be indicated as:
\begin{equation}\label{equ:path3}
\bm{p}_i= \sum ^{l_i-1}_{j=1} (\textbf{W}_{j}\bm{r}_j^{in} + \bm{b}_{j}),
\end{equation}
which utilizes a convolution neural network (CNN) to 
aggregate the relational path, considering the relation sequence. $l_i$ is the 
number of relations in $p_i$, and $\bm{r}_j^{in}$ refers to the $j$-th window of the relation sequence. For 
the special condition when $l_i=1$, we set the representation of the only relation to be $\bm{p}_i$.
$\textbf{W}_{j}$ is the convolution kernel and $\bm{b}_{j}$ is the optional bias.
The attention weight $\beta_i$ can be regarded as the confidence of corresponding rule for inference in $G$, which 
comes from the value of semantic similarity:
\begin{equation}
\label{equ:path2}\beta_i = \text{softmax}(\bm{p}_i, \bm{r}_{\scriptscriptstyle T}) 
= \frac {\text{exp}({\bm{p}_i}^\top \bm{r}_{\scriptscriptstyle T})}{\begin{matrix} \sum_{p_k \in \mathcal{P}_{h \to t}} 
\text{exp}({\bm{p}_k}^\top \bm{r}_{\scriptscriptstyle T}) \end{matrix}},
\end{equation}
in which $p_k$ refers to each single relational path in $\mathcal{P}_{h \to t}$. For further contrastive learning, 
representations of original sample, positive and negative path are $\bm{r}_{\scriptscriptstyle T}$, $\bm{p}_{h \to 
t}^{+}$ and $\bm{p}_{h \to t}^{-}$ respectively, in which $\bm{p}_{h \to t}^{+}$ and $\bm{p}_{h \to t}^{-}$ can be 
acquired with representations of $p^{+}_i$ and $p^{-}_i$.

\textbf{Rules and Interpretability.} RPC-IR extracts relational paths to capture the entity-independent information 
during the reasoning process, which can be treated as first-order rules in KGs. After training, RPC-IR obtains 
relational paths with attention weights, that are actually rules with confidence values extracted from the KG. For 
example, 
considering the target relation 
$\texttt{r}_{\scriptscriptstyle T}$, if the calculated attention weight of the relational path $(\texttt{r}_1, 
\texttt{r}_2,\dots,\texttt{r}_n)$ is $\beta$, then the 
structure and confidence 
$\beta \in [0,1]$ of 
the rule are derived simultaneously:
\begin{equation}\label{equ:rule2}
\begin{small}
\beta \; \texttt{r}_{\scriptscriptstyle T}(X, Y) \gets \texttt{r}_1(X, Z_1) \land \texttt{r}_2(Z_1, Z_2) \land 
\texttt{r}_n(Z_{n-1}, Y).
\end{small}
\end{equation}
During inference, variables $X, Z_1, Z_2,\dots, Z_n, Y$ are instantiated to entities $x, z_1, z_2, \dots, z_n, y$, and 
RPC-IR returns a relation with a confidence value by . 
There would be several 
learned rules in a single subgraph that provide the interpretable process of reasoning in KGs.

\subsection{Joint Training Strategy}
\begin{figure}[t]
	\centering
	\includegraphics[scale=0.77]{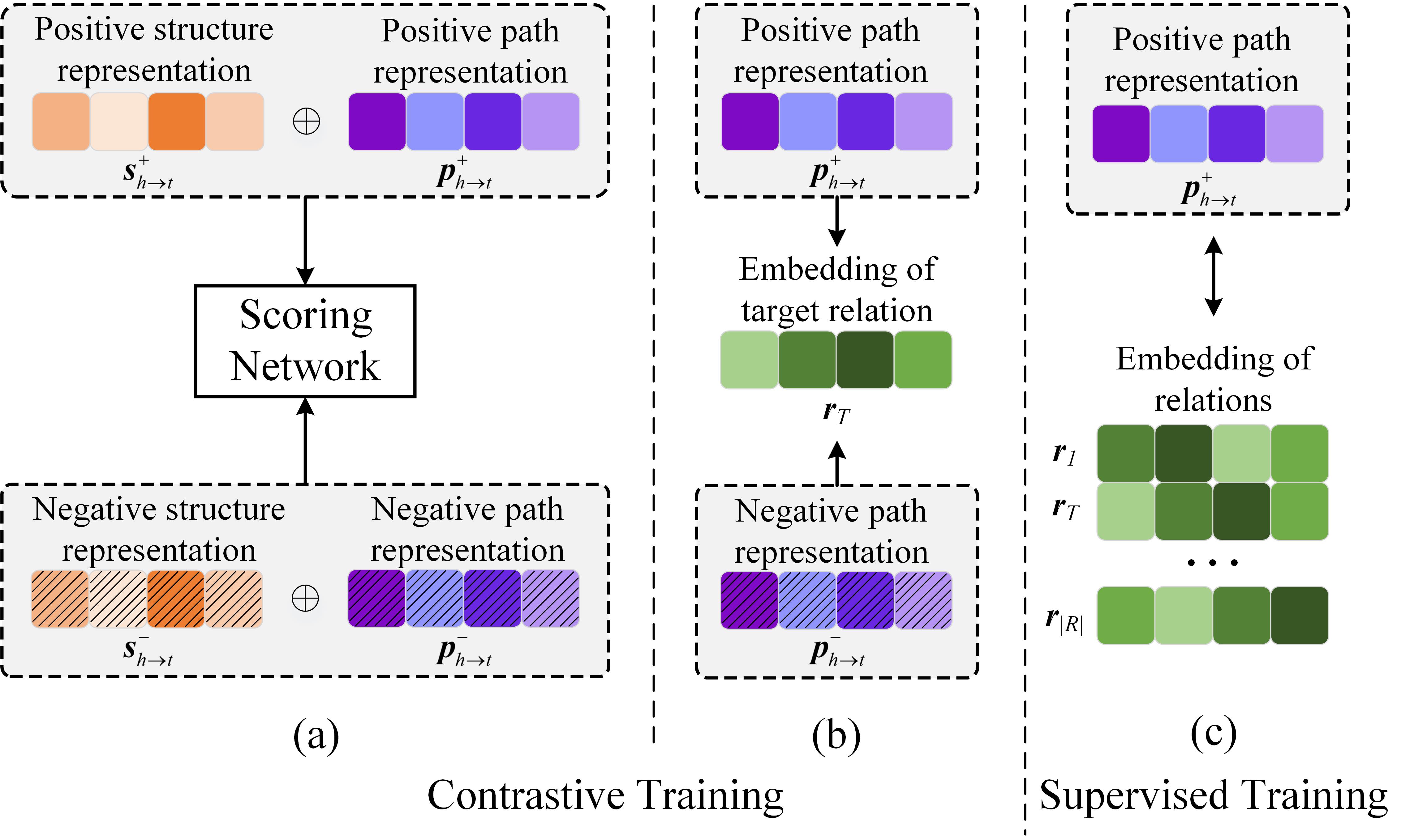}
	\caption{Joint Training Strategy.}
	\label{fig:3}
\end{figure}

In this step, we propose a joint training strategy combining the contrastive and supervised information. The 
contrastive training consists of the associative contrast and path contrast. The detailed descriptions are as follows:

\textbf{Associative Contrast.} 
In order to associate the topological structure of $\mathcal{G}_{T}$ denoted as $\bm{s}_{h \to t}$ and semantic 
information from the representation of paths denoted as 
$\bm{p}_{h \to t}$, we score the likelihood of target triple $e_{\scriptscriptstyle T}$ as:
\begin{equation}
\label{equ:score1} \bm{s}_{h \to t} = [\bm{z}_{\mathcal{G}_{T}}^{(L)} ; 
\bm{z}_{e_{\scriptscriptstyle T}}^{(L)}],
\end{equation}
\begin{equation}
\label{equ:score2} f(e_{\scriptscriptstyle T}, \mathcal{P}_{h \to t}, r_{\scriptscriptstyle T}) = 
\textbf{W}_s[\bm{s}_{h \to t}; \bm{r}_{\scriptscriptstyle T}; 
\bm{p}_{h \to t}],
\end{equation}
where $\textbf{W}_s$ is the weight matrix. $\bm{z}_{e_{\scriptscriptstyle T}}^{(L)}$ is the embedding concatenation of 
$(h, r_{\scriptscriptstyle T}, t)$ of all the $L$ layers' messages, which can be indicated as $\big[\bigoplus_{i=1}^L 
(\bm{z}^{(i)}_h ; \bm{z}^{(i)}_t)\big]$, 
where $\bigoplus$ is the concatenation operation.
$\bm{z}^{(L)}_{\mathcal{G}_{T}}$ refers 
to the global representation of $\mathcal{G}_{T}$, which is given by the average readout:
\begin{equation}\label{equ:readout}
\bm{z}^{(L)}_{\mathcal{G}_{T}} = \frac {1}{|\mathcal{V}_{T}|} \sum_{i \in 
	\mathcal{V}_{T}} \bm{z}^{(L)}_i,
\end{equation}
where $\mathcal{V}_{T}$ refers to the set of nodes in 
$\mathcal{G}_T$. 
We introduce margin-based loss to distance scores of positive and negative samples by an associative contrast:
\begin{equation}\label{equ:loss1}
\begin{aligned}
\mathcal{L}_G = 
&\sum _{e_{\scriptscriptstyle T} \in \mathcal{E}} \text{max}(0, \eta \, + \, f(e_{\scriptscriptstyle T}^{-}, 
\mathcal{P}_{h \to t}^{-}, r_{\scriptscriptstyle 
	T})\, \\
&- \, f(e_{\scriptscriptstyle T}^{+}, \mathcal{P}_{h \to t}^{+}, r_{\scriptscriptstyle T})).
\end{aligned}
\end{equation}
$e_{\scriptscriptstyle T}^{+}$ and $e_{\scriptscriptstyle T}^{-}$ refer to the positive and negative triple samples, 
where $e_{\scriptscriptstyle T}^{-}$ is the sample that replaces 
the 
head or tail of $e_{\scriptscriptstyle T}^{+}$. $\mathcal{E}$ is the set of all triples in $G$. The associative 
contrast loss is 
illustrated in Fig. \ref{fig:3}(a).

\textbf{Path Contrast.}
If we focus more on the semantic information given by relational paths, the contrastive learning should distinguish the 
target relation with negative 
paths and make it close to positive paths. Therefore, following the method of InfoNCE loss in 
\cite{DBLP:journals/corr/Oord} and assuming the 
samples are evenly distributed, the loss for path contrast is defined as:
\begin{equation}\label{equ:loss2}
\mathcal{L}_N = -\text{log} \Big[\frac {\text{exp}({\bm{p}_{h \to t}^+}^\top \bm{r}_{\scriptscriptstyle 
T})}{\text{exp}({\bm{p}_{h \to t}^+}^\top \bm{r}_{\scriptscriptstyle T})+\text{exp}({\bm{p}_{h \to t}^-}^\top 
\bm{r}_{\scriptscriptstyle T})}\Big].
\end{equation}
which is displayed in Fig. \ref{fig:3}(b).

\textbf{Supervised Training.}
Except for the contrastive learning, we implement the supervised prediction by computing the semantic similarity. With 
the 
representation of positive paths $\bm{p}^{+}_{h \to t}$ in $\mathcal{G}_T$, the supervised 
learning intends to compare it with the embedding of 
target relation $\bm{r}_{\scriptscriptstyle T}$. In our training strategy, we apply the cross entropy loss on all 
relation labels in 
$R$ to minimize the 
distance between 
$\bm{p}^{+}_{h \to t}$ and $\bm{r}_{\scriptscriptstyle T}$, and maximize the distances with other relations:
\begin{equation}\label{equ:loss3}
\mathcal{L}_C = -\text{log} \big[\frac {\text{exp}({\bm{p}_{h \to t}^+}^\top \bm{r}_{\scriptscriptstyle T})} 
{\begin{matrix} \sum_{r 
		\in R} 
	\end{matrix}\text{exp}({\bm{p}_{h \to t}^+}^\top \bm{r})}\big].
\end{equation} 
which is shown in Fig. \ref{fig:3}(c).

Eventually, the overall loss of our model is defined as the weighted summation of three losses, simultaneously 
optimizing them by a joint training 
process:
\begin{equation}\label{equ:loss4}
\mathcal{L} = \mathcal{L}_G + \lambda_1 \mathcal{L}_N + \lambda_2 \mathcal{L}_C,
\end{equation} 
where $\lambda_1$ and $\lambda_2$ are hyperparameters representing weights of path contrast loss and semantic 
similarity loss. 

\section{Experiments} \label{section_exp}
In this section, we firstly introduce benchmark datasets, experiment settings and details. Secondly, to verify the 
effectiveness of RPC-IR, we implement comparison experiments on relation prediction task. In addition, we use ablation 
studies, hyper-parameter sensitivity analysis and case studies to comprehensively demonstrate the performance.

\begin{table*}[h]
	\centering
	\caption{Comparison of AUC-PR (\%) results on inductive benchmarks derived from WN18RR, FB15K-237 and NELL-995. 
	$\dagger$ means we rerun the 
		project in the same device environment and record the results.}
	\renewcommand\arraystretch{1.3}
	\label{table:2}
	\begin{tabular}{ll|llll|llll|llll}
		\hline \hline
		\multirow{2}{*}{Category}                                                         
		& \multirow{2}{*}{Method} 
		& \multicolumn{4}{c|}{WN18RR}                                       
		& \multicolumn{4}{c|}{FB15K-237}                                    
		& \multicolumn{4}{c}{NELL-995} \\\cline{3-14}
		&                           
		& v1             & v2             & v3             & v4             
		& v1             & v2             & v3             & v4             
		& v1             & v2             & v3             & v4             \\ \hline
		\multirow{3}{*}{Rule-based}                                                   
		& RuleN		\cite{DBLP:conf/semweb/MeilickeFWRGS18}                  
		& 90.26          & 89.01          & 76.46          & 85.75          
		& 75.24          & 88.70          & 91.24          & 91.79          
		& 84.99          & 88.40          & 87.20          & 80.52          \\
		& Neural-LP	\cite{DBLP:conf/nips/YangYC17}               
		& 86.02          & 83.78          & 62.90          & 82.06          
		& 69.64          & 76.55          & 73.95          & 75.74          
		& 64.66          & 83.61          & 87.58          & 85.69          \\
		& DRUM		\cite{DBLP:conf/nips/SadeghianADW19}                 
		& 86.02          & 84.05          & 63.20          & 82.06          
		& 69.71          & 76.44          & 74.03          & 76.20          
		& 59.86          & 83.99          & 89.71          & 85.94          \\ \hline
		\multirow{2}{*}{\begin{tabular}[c]{@{}l@{}}Graph-based\end{tabular}} 
		& GraIL		\cite{DBLP:conf/icml/TeruDH20}                   
		& 94.32              & 94.18              & 85.80              & 92.72          
		& \underline{84.69}  & \underline{90.57}              & 91.68              & \underline{94.46}          
		& \underline{86.05}  & 92.62              & 93.34              & \underline{87.50}          \\
		& CoMPILE$\dagger$	\cite{DBLP:conf/aaai/MAI}                  
		& \underline{98.29}  & \underline{99.36}  & \underline{93.60}  & \textbf{99.51} 
		& 83.06              & 90.21              & \underline{93.12}  & 93.24          
		& 82.39              & \underline{93.30}  & \underline{95.71}  & 52.98         \\ \hline
		Ours
		& RPC-IR                      
		& \textbf{98.87}     & \textbf{99.41}     & \textbf{93.76}     & \underline{98.75}          
		& \textbf{87.24}     & \textbf{92.75}     & \textbf{93.93}     & \textbf{95.26} 
		& \textbf{88.12}     & \textbf{94.12}     & \textbf{96.10}     & \textbf{87.81} \\ \hline \hline
	\end{tabular}
\end{table*} 

\begin{table*}[h]
	\centering
	\caption{Comparison of Hits@10 (\%) results on inductive benchmarks derived from WN18RR, FB15K-237 and NELL-995.}
	\renewcommand\arraystretch{1.3}
	\label{table:3}
	\begin{tabular}{ll|llll|llll|llll}
		\hline \hline
		\multirow{2}{*}{Category}                                                         
		& \multirow{2}{*}{Method} 
		& \multicolumn{4}{c|}{WN18RR}                                       
		& \multicolumn{4}{c|}{FB15K-237}                                    
		& \multicolumn{4}{c}{NELL-995} \\\cline{3-14}
		&                           
		& v1             & v2             & v3             & v4             
		& v1             & v2             & v3             & v4             
		& v1             & v2             & v3             & v4             \\ \hline
		\multirow{3}{*}{Rule-based}                                                   
		& RuleN		\cite{DBLP:conf/semweb/MeilickeFWRGS18}                  
		& 80.85          & 78.23          & 53.39          & 71.59          
		& 49.76          & 77.82          & \textbf{87.69} & 85.60          
		& 53.50          & 81.75          & 77.26          & 61.35          \\
		& Neural-LP	\cite{DBLP:conf/nips/YangYC17}               
		& 74.37          & 68.93          & 46.18          & 67.13          
		& 52.92          & 58.94          & 52.90          & 55.88          
		& 40.78          & 78.73          & 82.71          & 80.58 \\
		& DRUM		\cite{DBLP:conf/nips/SadeghianADW19}                 
		& 74.37          & 68.93          & 46.18          & 67.13          
		& 52.92          & 58.73          & 52.90          & 55.88          
		& 19.42          & 78.55          & 82.71          & \textbf{80.58} \\ \hline
		\multirow{2}{*}{\begin{tabular}[c]{@{}l@{}}Graph-based\end{tabular}} 
		& GraIL		\cite{DBLP:conf/icml/TeruDH20}                   
		& \underline{82.45} & \underline{78.68}  & 58.43              & \underline{73.41}          
		& \underline{64.15} & 81.80              & 82.83              & \textbf{89.29}          
		& \underline{59.50} & \underline{93.25}  & 91.41              & \underline{73.19}          \\
		& CoMPILE$\dagger$	\cite{DBLP:conf/aaai/MAI}                  
		& 81.91 & 76.64  & \underline{60.69}  & 71.80
		& 62.20             & \underline{82.01}  & \underline{84.67}  & 87.44          
		& 58.33             & 88.86              & \underline{93.63}  & 60.81          \\ \hline
		Ours
		& RPC-IR                      
		& \textbf{85.11}    & \textbf{81.63}     & \textbf{62.40}     & \textbf{76.35}          
		& \textbf{67.56}    & \textbf{82.53}     & 84.39 		      & \underline{89.22} 
		& \textbf{59.75}    & \textbf{93.28}     & \textbf{94.01}     & 71.82 \\ \hline \hline
	\end{tabular}
\end{table*} 

\subsection{Experimental Settings} 
\textbf{Datasets.}
The inductive link prediction datasets \cite{DBLP:conf/icml/TeruDH20} are derived from WN18RR 
\cite{DBLP:conf/aaai/DettmersMS018}, FB15K-237 \cite{DBLP:conf/emnlp/ToutanovaCPPCG15} and NELL-995 
\cite{DBLP:conf/emnlp/XiongHW17}, and have been generated into four versions respectively. In each dataset, 
there is no intersection between entities in the train set and test set for the fully-inductive setting. The statistics 
of benchmark datasets are illustrated in TABLE \ref{table:1}. In particular, each version of a dataset consists of a 
pair of knowledge graphs, \textit{train} and \textit{ind-test}, whose entities are totally different. Meanwhile, the 
knowledge graph in \textit{train} contains all the relations in \textit{ind-test}.

\textbf{Metrics.}
We demonstrate the effectiveness of RPC-IR by comparing it with other methods on inductive relation prediction tasks. 
In the comparison, we implement both classification and ranking metrics to evaluate the model.

AUC-PR is an indicator for classification task computing the area under prediction-recall curve. 
In order to calculate the AUC-PR, we apply the scores considering the subgraph and paths on positive and negative 
samples. 

For the ranking metric Hits@10, we evaluate it in a general mode by ranking the test triples among 50 randomly negative 
samples. We record the mean results over multiple runs considering the random seeds and samples. 

\begin{table}[t]
	\centering
	\caption{Statistics of Datasets.}
	\scriptsize
	\renewcommand\arraystretch{1.3}
	\label{table:1}
	\begin{tabular}{p{1mm}p{7.5mm}p{2mm}p{4mm}p{5mm}p{2mm}p{4mm}p{5mm}p{2mm}p{4mm}p{5mm}}
		\hline \hline
		&          		& \multicolumn{3}{c}{WN18RR} & \multicolumn{3}{c}{FB15K-237} & \multicolumn{3}{c}{NELL-995} \\
		&          		& \#R    & \#E     & \#Tr    & \#R     & \#E      & \#Tr     & \#R    & \#E      & \#Tr     \\ 
		\hline
		\multirow{2}{*}{v1} & \textit{train}    		
		& 9      & 2,746    & 6,678    & 183     & 2,000     & 5,226     & 14     & 10,915    & 5,540     \\
		& \textit{ind-test}		
		& 9      & 922      & 1,991    & 146     & 1,500     & 2,404     & 14     & 225       & 1,034      \\ \hline
		\multirow{2}{*}{v2} & \textit{train}    		
		& 10     & 6,954    & 18,968   & 203     & 3,000     & 12,085    & 88     & 2,564     & 10,109    \\
		& \textit{ind-test} 		
		& 10     & 2,923    & 4,863    & 176     & 2,000     & 5,092     & 79     & 4,937     & 5,521     \\ \hline
		\multirow{2}{*}{v3} & \textit{train}   		
		& 11     & 12,078   & 32,150   & 218     & 4,000     & 22,394    & 142    & 4,647     & 20,117    \\
		& \textit{ind-test} 		
		& 11     & 5,084    & 7,470    & 187     & 3,000     & 9,137     & 122    & 4,921     & 9,668     \\ \hline
		\multirow{2}{*}{v4} & \textit{train}    		
		& 9      & 3,861    & 9,842    & 222     & 5,000     & 33,916    & 77     & 2,092     & 9,289     \\
		& \textit{ind-test} 		
		& 9      & 7,208    & 15,157   & 204     & 3,500     & 14,554    & 61     & 3,294     & 9,520     \\ \hline
		\hline
	\end{tabular}
\end{table}

\begin{table*}[h]
	\centering
	\caption{Ablation results of AUC-PR on inductive benchmarks derived from WN18RR, FB15K-237 and NELL-995.}
	\renewcommand\arraystretch{1.3}
	\label{table:4}
	\begin{tabular}{l|p{6mm}p{6mm}p{6mm}p{6mm}|p{6mm}|p{6mm}p{6mm}p{6mm}p{6mm}|p{6mm}|p{6mm}p{6mm}p{6mm}p{6mm}|p{6mm}}
		\hline \hline
		\multirow{2}{*}{Method} 
		& \multicolumn{5}{c|}{WN18RR}                                       
		& \multicolumn{5}{c|}{FB15K-237}                                    
		& \multicolumn{5}{c}{NELL-995}                                      \\\cline{2-16}
		& v1             & v2             & v3             & v4             & Avg   
		& v1             & v2             & v3             & v4             & Avg   
		& v1             & v2             & v3             & v4             & Avg   \\ \hline
		RPC-IR w/o paths            
		& 93.04          & 95.15          & 87.48          & 94.22          & 92.47   
		& 84.56          & 91.23          & 91.97          & 92.90          & 90.17   
		& 81.94          & 91.59          & 89.90          & 73.81          & 84.31   \\
		$\Delta$
		& $\downarrow$5.83           & $\downarrow$4.26           & $\downarrow$6.28           & 
		$\downarrow$4.53        & $\downarrow$5.23  
		& $\downarrow$2.68           & $\downarrow$1.52           & $\downarrow$1.96           & 
		$\downarrow$2.36        & $\downarrow$2.13 
		& $\downarrow$6.18           & $\downarrow$2.53           & $\downarrow$6.20           & 
		$\downarrow$14.00       & $\downarrow$7.23 \\ 
		\hline
		RPC-IR w/o contrasts         
		& 95.38          & 95.26          & 87.90          & 95.05          & 93.40   
		& 83.71          & 91.69          & 93.89          & 91.85          & 90.29   
		& 82.97          & 90.83          & 91.52          & 78.45          & 85.94   \\
		$\Delta$
		& $\downarrow$3.49           & $\downarrow$4.15           & $\downarrow$5.86           & 
		$\downarrow$3.70        & $\downarrow$4.30
		& $\downarrow$3.53           & $\downarrow$1.06           & $\downarrow$0.04           & 
		$\downarrow$3.41        & $\downarrow$2.01 
		& $\downarrow$5.15           & $\downarrow$3.29           & $\downarrow$4.58           & 
		$\downarrow$9.36        & $\downarrow$5.60 \\ \hline
		RPC-IR                      
		& \textbf{98.97} & \textbf{99.41} & \textbf{93.76} & \textbf{98.75} & \textbf{97.70}
		& \textbf{87.24} & \textbf{92.75} & \textbf{93.93} & \textbf{95.26} & \textbf{92.30}
		& \textbf{88.12} & \textbf{94.12} & \textbf{96.10} & \textbf{87.81} & \textbf{91.54}  
		\\ 
		\hline \hline
	\end{tabular}
\end{table*}

\textbf{Experimental Details.}
For the subgraph extraction, we obtain 3-hop enclosing subgraphs by the double vertex labeling. In the graph embedding 
process, we employ a 3-layer GCN with the dimension of the relations and entities as 32. The 
dropout rate in triples of subgraphs is set to 0.5. When extracting relational paths, we use the max length $L_{max}=3$ 
for WN18RR and FB15K-237, and $L_{max}=2$ for NELL-995 considering the high time complexity. In order to generate 
negative paths, we randomly replace a relation in each relational path. During the training process, the batch size is 
set as 16 and we use Adam \cite{DBLP:journals/corr/KingmaB14} as optimizer with learning rate being 0.001.

\textbf{Baselines.}
The baselines for comparison are previous methods for inductive reasoning in KGs. RuleN 
\cite{DBLP:conf/semweb/MeilickeFWRGS18} is the statistical rule-based inductive method which obtains outstanding 
performance in the transductive setting. Neural-LP \cite{DBLP:conf/nips/YangYC17} and DRUM 
\cite{DBLP:conf/nips/SadeghianADW19} are differentiable methods, which generate rules during the reasoning process. 
Graph-based inductive methods GraIL \cite{DBLP:conf/icml/TeruDH20} and CoMPILE \cite{DBLP:conf/aaai/MAI} can implement 
inductive reasoning as well, but they implement the prediction without interpretability by explicit rules.
As a note, to reduce the influence of experimental environment and implement further comparison, we rerun 
state-of-the-art method CoMPILE \cite{DBLP:conf/aaai/MAI} with corresponding settings from the original project and 
record the results. Other results of baselines are from the comparison results in \cite{DBLP:conf/icml/TeruDH20}. 

\begin{figure*}[h]
	\centering
	\includegraphics[scale=0.4]{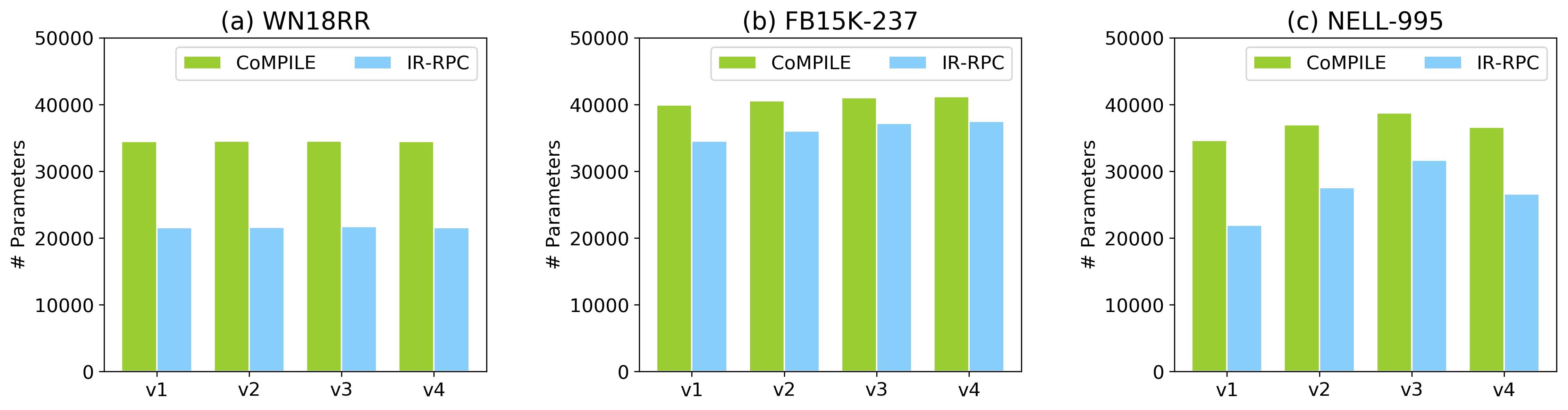}
	\caption{Comparing numbers of parameters of RPC-IR and state-of-the-art method CoMPILE.}
	\label{fig:5}
\end{figure*}

\subsection{Comparison Results}
\textbf{Comparison of Prediction Results.}
TABLE \ref{table:2} and \ref{table:3} show the comparison results of relation prediction.
Compared with the listed baselines, RPC-IR significantly outperforms them among the vast majority of datasets in two 
metrics. The detailed analysis is as follows:
\begin{itemize}
	\item For the rule-based inductive methods, the average boosts of RPC-IR on WN18RR, FB15K-237 and NELL-995 in 
	AUC-PR are 12.53\%, 6.42\% and 7.04\% respectively compared with the rule-based inductive method RuleN 
	\cite{DBLP:conf/ijcai/MeilickeCRS19}. 
	RPC-IR is also superior to differentiable rule-based inductive methods Neural-LP and DRUM in terms of the 
	classification performance. On the ranking task, RPC-IR outperforms other rule-based inductive methods on most 
	datasets, except for Hits@10 results on FB15K-237\_v3 and NELL-995\_v4. We attribute this phenomenon to the general 
	performance of graph-based pattern.  
	\item After observation, graph-based inductive methods are generally more effective on most datasets than 
	rule-based inductive 
	methods. Comparing with more competitive graph-based inductive methods, RPC-IR owns optimal results among these 
	datasets 
	in two metrics, which illustrate its superiority as well. RPC-IR results in as much as 6.15\%, 2.82\% and 2.44\% 
	average performance improvements in AUC-PR comparing to GraIL, which is the basic graph-based inductive method. 
	Especially on WN18RR, the classification performance is superior to GraIL, reflecting that the relational path 
	contrast strategy is effective on the more sparse dataset shown in TABLE \ref{table:intro}.
	As for the state-of-the-art method CoMPILE, our method performs better on most datasets in terms of both metrics in 
	the same experimental environment. 
\end{itemize}

\textbf{Comparison of Complexity.}
Moreover, RPC-IR needs less parameters than the state-of-the-art CoMPILE when training the model, which means we 
achieve lower model complexity. The results on three datasets are shown in Fig. \ref{fig:5}. (a), (b) and (c) 
severally, in 
which the green bars refer to the numbers of parameters of CoMPILE, and the blue bars refer to these of RPC-IR. 
Although RPC-IR owns slightly lower results on a few datasets than CoMPILE, the complexity of RPC-IR is evidently lower 
than that of CoMPILE on all datasets. For example, CoMPILE gets better AUC-PR value on WN18RR\_v4, but owns 34,465 
parameters while the number of parameters of RPC-IR is 21,536, reflecting the performance superiority of RPC-IR on 
WN18RR\_v4 from an aspect. 

\subsection{Ablation Results}
In this subsection, we intend to investigate impacts of relational paths and contrasts in inductive learning 
respectively. TABLE \ref{table:4} indicates the results when training the model by RPC-IR without these factors on all 
three datasets. For the relational paths, we remove it from our method to verify their contributions and call it 
``\textbf{RPC-IR w/o paths}". The same operation is implemented to the contrasts and the method is called 
``\textbf{RPC-IR w/o contrasts}". Because of the fair comparison, other parameters remain the same during training and 
testing.

From TABLE \ref{table:4}, we can easily figure that the reduction of AUC-PR values occurs when we train RPC-IR without 
relational paths and contrastive learning. After removing relational paths, the average AUC-PR values on WN18RR, 
FB15K-237 and 
NELL-995 reduce by 5.43\%, 3.00\% and 8.01\% severally. The lack of contrasts results in corresponding reductions 
by 4.50\%, 2.88\% and 6.38\% on three datasets. We can also observe that relational paths contribute more in inductive 
reasoning than contrasts, but better results are obtained by adding relational paths and contrasts simultaneously. 
In addition, by observing results after removing two factors, it shows that relational paths and contrasts are more 
effective on NELL-995 than other two datasets. 

\begin{table*}[h]
	\centering
	\scriptsize
	\caption{Rules derived from three versions of datasets WN18RR\_v1, FB15K-237\_v1 and NELL-995\_v1.}
	\renewcommand\arraystretch{1.3}
	\label{table:5}
	\begin{tabular}{rl}
		\hline
		\multicolumn{2}{c}{WN18RR\_v1} \\ \hline
		0.99 & 
		\texttt{hypernym}$(X,Y)\gets$
		\texttt{verb\_group}$(X,Z_1)\,\land$
		\texttt{hypernym}$(Z_1,Z_2)\,\land$
		\texttt{hypernym}$(Z_2,Y)$ \\ 
		0.50 &
		\texttt{hypernym}$(X,Y)\gets$
		\texttt{derivationally\_related\_form}$(X,Z_1)\,\land$
		\texttt{derivationally\_related\_form}$(Z_1,Z_2)\,\land$
		\texttt{hypernym}$(Z_2,Y)$ \\ 
		\textcolor[RGB]{202,12,22}{\textless{}0.01} &
		\textcolor[RGB]{202,12,22}{\texttt{hypernym}$(X,Y)\gets\,$
			\texttt{derivationally\_related\_form}$(X,Z)\,\land$
			\texttt{derivationally\_related\_form}$(Z,Y)$} \\ 
		\hline
		
		0.41 & 
		\texttt{verb\_group}$(X,Y)\gets$
		\texttt{derivationally\_related\_form}$(X,Z_1)\,\land$
		\texttt{derivationally\_related\_form}$(Z_1,Z_2)\,\land$
		\texttt{verb\_group}$(Z_2,Y)$ \\ 
		0.41 & 
		\texttt{verb\_group}$(X,Y)\gets$
		\texttt{verb\_group}$(X,Z_1)\,\land$
		\texttt{derivationally\_related\_form}$(Z_1,Z_2)\,\land$
		\texttt{derivationally\_related\_form}$(Z_2,Y)$ \\
		0.18 &
		\texttt{verb\_group}$(X,Y)\gets$
		\texttt{verb\_group}$(X,Z)\,\land$
		\texttt{verb\_group}$(Z,Y)$ \\
		\textcolor[RGB]{202,12,22}{\textless{}0.01} &
		\textcolor[RGB]{202,12,22}{\texttt{verb\_group}$(X,Y)\gets\,$
			\texttt{hypernym}$(X,Y)$} \\ 
		\hline
		\multicolumn{2}{c}{FB15K-237\_v1} \\ \hline
		1.00 &  
		\texttt{location/contains}$(X,Y)\gets\,$
		\texttt{location/contains}$(X,Z_1)\,\land$
		\texttt{location/state}$(Z_1,Z_2)\,\land$
		\texttt{location/contains}$(Z_2,Y)$ \\ 
		1.00 &  
		\texttt{location/contains}$(X,Y)\gets$
		\texttt{location/contains}$(X,Z)\,\land$
		\texttt{location/contains}$(Z,Y)$ \\  
		0.45 &
		\texttt{location/contains}$(X,Y)\gets$
		\texttt{location/contains}$(X,Z)\,\land$
		\texttt{location/adjoins}$(Z,Y)$ \\
		\textcolor[RGB]{202,12,22}{\textless{}0.01} &
		\textcolor[RGB]{202,12,22}{\texttt{location/contains}$(X,Y)\gets\,$
			\texttt{gardening\_hint/split\_to}$(X,Y)$} \\ 
		\hline
		
		
		0.99 &  
		\texttt{person/religion}$(X,Y)\gets$
		\texttt{friendship/participant}$(X,Z)\,\land$
		\texttt{person/religion}$(Z,Y)$ \\  
		0.82 &
		\texttt{person/religion}$(X,Y)\gets$
		\texttt{dated/participant}$(X,Z)\,\land$
		\texttt{person/religion}$(Z,Y)$ \\
		\textcolor[RGB]{202,12,22}{\textless{}0.01} &
		\textcolor[RGB]{202,12,22}{\texttt{person/religion}$(X,Y)\gets\,$
			\texttt{marriage/type\_of\_union}$(X,Z_1)\,\land$
			\texttt{/location\_of\_ceremony}$(Z_1,Z_2)\,\land$
			\texttt{location/religion}$(Z_2,Y)$} \\
		\hline
		\multicolumn{2}{c}{NELL-995\_v1} \\ \hline
		1.00 &  
		\texttt{subpartOf}$(X,Y)\gets$
		\texttt{subpartOf}$(X,Z)\,\land$
		\texttt{subpartOf}$(Z,Y)$ \\
		0.97 &
		\texttt{subpartOf}$(X,Y)\gets$
		\texttt{agentBelongsToOrganization}$(X,Z_1)\,\land$
		\texttt{agentBelongsToOrganization}$(Z_1,Z_2)\,\land$
		\texttt{subpartOf}$(Z_2,Y)$ \\ 
		0.38 &
		\texttt{subpartOf}$(X,Y)\gets$
		\texttt{agentBelongsToOrganization}$(X,Z)\,\land$
		\texttt{subpartOf}$(Z,Y)$ \\
		\textcolor[RGB]{202,12,22}{\textless{}0.01} &
		\textcolor[RGB]{202,12,22}{\texttt{subpartOf}$(X,Y)\gets\,$
			\texttt{agentcollaborateswithagent}$(X,Y)$} \\
		\hline
		
		1.00 & 
		\texttt{worksFor}$(X,Y)\gets$
		\texttt{agentControls}$(X,Z_1)\,\land$
		\texttt{agentCollaboratesWithAgent}$(Z_1,Z_2)\,\land$
		\texttt{worksFor}$(Z_2,Y)$ \\ 
		0.53 &
		\texttt{worksFor}$(X,Y)\gets$
		\texttt{worksFor}$(X,Z)\,\land$
		\texttt{subpartOfOrganization}$(Z,Y)$ \\	
		\textcolor[RGB]{202,12,22}{\textless{}0.01} &
		\textcolor[RGB]{202,12,22}{\texttt{worksfor}$(X,Y)\gets\,$
			\texttt{topmemberoforganization}$(X,Y)$} \\ 	
		\hline
	\end{tabular}
\end{table*} 

\subsection{Hyper-parameter Sensitivity Analysis}
In this subsection, we analyze the sensitivity of hyper-parameters on different datasets.
In our model, $\lambda_1$ and $\lambda_2$ are critical for adjusting functions of the supervised and self-supervised 
learning during training, so we rerun the training process in different values of $\lambda_1$ and $\lambda_2 \in [0.2, 
1.2]$, and record the AUC-PR results to analysis the effectiveness of them. Two versions of datasets, WN18RR\_v1 and 
FB15K-237\_v1, are utilized to help achieve the analysis. Considering the training time, we run 150 epochs on 
WN18RR\_v1 and 60 epochs on FB15K-237\_v1, and show the test results in of Fig. \ref{fig:5} (a) and (b) respectively. 
The test results written on the heat maps are mean values after 5 runs.

From the distribution of mean results, we get several observation. Firstly, for WN18RR\_v1, better results gather at 
the lower right corner of the heat map, especially when $\lambda_1=1.0$ and $\lambda_2=1.2$. Secondly, for 
FB15K-237\_v1, apparently the best results distribute near the diagonal, which means the supervised 
and self-supervised learning are equally important for inductive reasoning. When $\lambda_1=0.8$ and $\lambda_2=0.8$, 
RPC-IR obtains the best test result. The distributions are distinct on different datasets, and it might be related to 
the number of paths in a subgraph, for the average number paths in an enclosing subgraph of WN18RR are less than that 
of FB15K-237 shown in Fig. \ref{table:intro}.

\begin{figure}[t]
	\centering
	\includegraphics[scale=0.31]{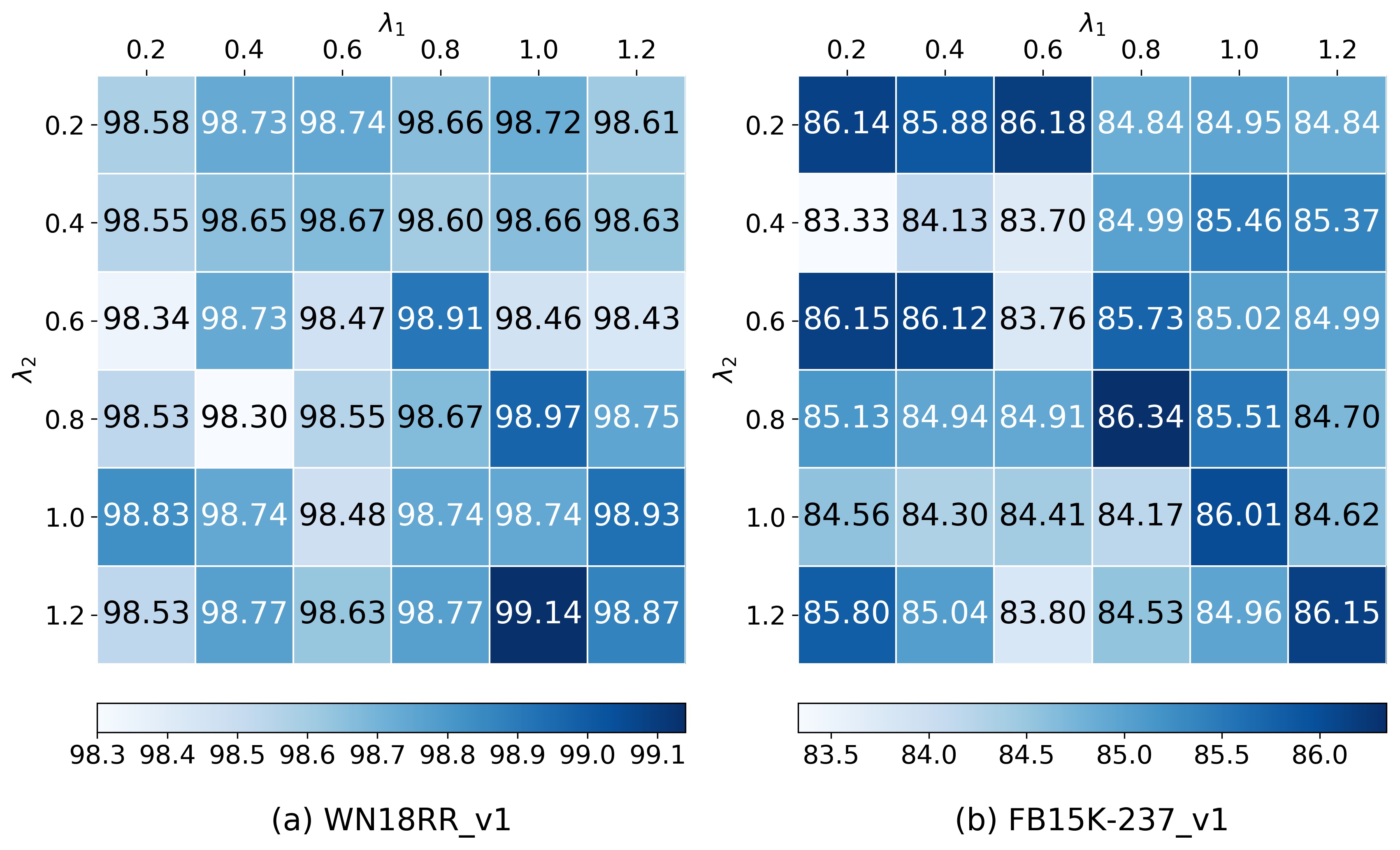}
	\caption{Effectiveness evaluation by AUC-PR of parameters $\lambda_1$ and $\lambda_2$ over two datasets.}
	\label{fig:4}
\end{figure}

\subsection{Case Studies}
As stated in section \ref{section_method}, a crucial advantage of RPC-IR in inductive learning is to represent 
first-order rules for reasoning explicitly. 
TABLE \ref{table:5} shows examples of derived rules by RPC-IR on the first version of WN18RR, FB15K-237 and NELL-995. 
The value in front of each rule is the confidence value in the corresponding subgraph. Rules in the same block are with 
the same head, which is generalized from the target triple, 
and the body is generalized from the reasoning path when predicting. 
The rules in \textcolor[RGB]{202,12,22}{red 
	text} are with the weight less than 0.01, which is unreasonable when inference.
Overall, RPC-IR implements the interpretability by these explicit rules. 

\section{Conclusion} \label{section_con}
We propose a novel inductive reasoning and rule learning approach by relational path contrast in KGs, named RPC-IR. To 
acquire the entity independence semantics from latent rules and solve the deficient supervision in a single subgraph, 
RPC-IR extracts relational paths in each subgraph and introduces contrastive learning to obtain self-supervised 
information. 
The experiments on three fully-inductive datasets show the effectiveness of IR-RPC, and comprehensively demonstrate the 
impacts for relational paths and contrasts. 

RPC-IR still needs improving on scalability and performance. In the future, we intend to implement inductive 
reasoning and rule learning on more datasets, for example the KGs of curriculum areas, or commonsense knowledge 
graphs whose entities are free-form texts.

\section*{Acknowledgment}

The authors would like to thank...

\ifCLASSOPTIONcaptionsoff
  \newpage
\fi



%
\bibliographystyle{IEEEtran}
\bibliography{reference}

%

%
%
%




\end{document}